\documentclass{ws-p9-75x6-50}

\begin{document}

\title{Pairing correlations in nuclear systems, from infinite matter to 
       finite nuclei}

\author{\O.\ ELGAR\O Y, T.\ ENGELAND, M.\ HJORTH-JENSEN AND E.\ OSNES}

\address{Department of Physics, University og Oslo, N-0316 Oslo, Norway}

\maketitle

\abstracts{Finite nuclei such as those found  
in the chain
of even tin isotopes from 
$^{102}$Sn to $^{130}$Sn, exhibit a near constancy of the 
$2^+_1-0^+_1$ excitation energy, a constancy which can be related
to strong pairing correlations and the near degeneracy in energy 
of the relevant single particle orbits.
Large shell-model calculations for these isotopes reveal that 
the major contribution to pairing correlations in the tin isotopes stems
from the $^1S_0$ partial wave in the nucleon-nucleon interaction.
Omitting this partial wave and the $^3P_2$ wave
in the construction of an effective interaction, 
results in a spectrum which has essentially no correspondence with experiment. 
These partial wave are also of importance for infinite neutron matter and
nuclear matter
and give the largest contribution to the pairing 
interaction and energy gap in neutron star matter. }

\section{Introduction}

One of the fundamental, yet unsolved problems of nuclear theory is to describe the properties of
 complex nuclei and infinite nuclear matter
in terms of their constituent particles and the interaction among them.
There are two major obstacles to the solution of this problem. Firstly, we are dealing
with a quantal many-body problem which cannot be solved exactly. Secondly, the basic
nucleon-nucleon (NN) interaction is not well known. Thus, it may be difficult to know 
whether an eventual failure to solve this problem should be ascribed to the many-body 
methods or the interaction model used. On the other hand, these two uncertainties are 
intimately connected. In any model chosen to approximate the original many-body problem
one has to apply an interaction which is consistent with the particular degrees of freedom 
considered. This amounts to correcting the original interaction for the degrees of freedom 
not explicitly included in the many-body treatment, thus yielding a so-called effective interaction.

In principle, one should start from an NN interaction derived from the interaction between quarks. 
Although attempted, this program has not been quantitatively successful. Thus, one has to be 
content with using an NN interaction derived from meson-exchange models which reproduce the 
relevant two-nucleon data. Such interactions are generally termed realistic interactions. 
Once the basic NN interaction has been established, it should be employed in a quantal 
many-body approximation to the nuclear structure problem of interest. 
One such approximation 
is the spherical shell model, which has provided a successful microscopic approach for nuclei 
near closed shells. In shell-model calculations one typically employs an effective interaction
taylored to include those degrees of freedom which are thought to be relevant. 

Since the effective interactions employed in shell-model calculations 
are always the outcome of some truncations in the many-body expansion,
the shell model may then provide a useful testing ground for 
the various approximations made. Furthermore, the shell-model wave function
can be used to extract information on specific correlations in nuclei, such as
pairing correlations. 

In this study we focus on the link between different partial waves in the
NN interaction and the results of large-scale shell-model calculations
in the Sn isotopes. We 
demonstrate that those partial waves which are important for the 
presence of superfluidity in infinite neutron star matter, play a crucial role in describing
the  near constancy of the 
$2^+_1-0^+_1$ excitation energy in the chain
of even tin isotopes from 
$^{102}$Sn to $^{130}$Sn. This near constancy    
is in turn related to strong pairing correlations.  

This contribution falls in four sections. After the above introductory words, we give a 
brief review of pairing in infinite neutron matter. Shell-model
analyses of different approaches to the effective interaction are in turn made in section
\ref{sec:sec3} and concluding remarks 
are given in section \ref{sec:sec4}.
\section{Pairing in infinite neutron matter}\label{sec:sec2}

The presence of neutron superfluidity in 
the crust and the inner part 
of neutron stars 
are considered well established 
in the physics of these compact stellar objects. 
In the low density outer part of a neutron star, 
the neutron superfluidity is expected 
mainly in the attractive $^1S_0$ channel. 
At higher density, the nuclei in the crust dissolve, and one 
expects a region consisting of a quantum liquid of neutrons and 
protons in beta equilibrium. 
The proton contaminant should be superfluid 
in the $^1S_0$ channel, while neutron superfluidity is expected to  
occur mainly in the coupled $^3P_2$-$^3F_2$ two-neutron channel. 
In the core of the star any superfluid 
phase should finally disappear.
 
The presence of two different superfluid regimes 
is suggested by the known trend of the 
nucleon-nucleon (NN) phase shifts 
in each scattering channel. 
In both the $^1S_0$ and $^3P_2$-$^3F_2$ channels the
phase shifts indicate that the NN interaction is attractive. 
In particular for the $^1S_0$ channel, the occurrence of 
the well known virtual state in the neutron-neutron channel
strongly suggests the possibility of a 
pairing condensate at low density, 
while for the $^3P_2$-$^3F_2$ channel the 
interaction becomes strongly attractive only
at higher energy, which therefore suggests a possible 
pairing condensate
in this channel at higher densities. 
In recent years the BCS gap equation
has been solved with realistic interactions, 
and the results confirm
these expectations. 

The $^1S_0$ neutron superfluid is relevant for phenomena
that can occur in the inner crust of neutron stars, like the 
formation of glitches, which may to be related to vortex pinning  
of the superfluid phase in the solid crust \cite{glitch}. 
The results of different groups are in close agreement
on the $^1S_0$ pairing gap values and on 
its density dependence, which
shows a peak value of about 3 MeV at a Fermi momentum close to
$k_F \approx 0.8\; {\rm fm}^{-1}$ \cite{bcll90,kkc96,eh98,sclbl96}. 
All these calculations adopt the bare
NN interaction as the pairing force, and it has been pointed out
that the screening by the medium of the interaction 
could strongly reduce
the pairing strength in this channel \cite{sclbl96,chen86,ains89}. 
However, the issue of the 
many-body calculation of the pairing 
effective interaction is a complex
one and still far from a satisfactory solution.

The precise knowledge of the $^3P_2$-$^3F_2$ pairing gap is of 
paramount relevance for, e.g.\  the cooling of neutron stars, 
and different values correspond to drastically
different scenarios for the cooling process.
Generally, the gap suppresses the cooling by a factor
$\sim\exp(-\Delta/T)$ (where $\Delta$ is the energy gap)
which is severe for
temperatures well below the gap energy.
Unfortunately, only few and partly
contradictory calculations of the pairing gap exist in the literature, 
even at the level of the bare NN interaction 
\cite{amu85,bcll92,taka93,elga96,khodel97}. 
However, when comparing the results, one should note that the  
NN interactions used in these calculations are not phase-shift 
equivalent, i.e.\  they do not 
predict exactly the same NN phase shifts.  
Furthermore, for the interactions used in 
Refs.~\cite{amu85,bcll92,taka93,elga96} the predicted 
phase shifts do not agree accurately with modern phase shift 
analyses, and the fit of the NN data has typically 
$\chi^2/{\rm datum}\approx 3$.  
Progress has 
however been made not only in the accuracy and the consistency of the 
phase-shift analysis, but also in the fit of realistic NN interactions 
to these data.  As a result, several new NN interactions have 
been constructed which fit the world data for $pp$ and $np$ scattering 
below 350 MeV with high precision.  Potentials like the recent 
Argonne $V_{18}$ \cite{v18}, the CD-Bonn \cite{cdbonn} 
or the new Nijmegen potentials \cite{nim} yield a 
$\chi^2/{\rm datum}$ of about 1 and may be called phase-shift 
equivalent.  
In Table \ref{tab:pgaps} we show the recent non-relativistic
pairing gaps for the $^3P_2$-$^3F_2$ partial waves, where
effective nucleon masses from the  lowest-order Brueckner-Hartree-Fock
calculation have been 
employed, see Ref.\ \cite{beehs98} for more details.
These results are for pure neutron matter and we observe that
up to $k_F\sim 2$ fm$^{-1}$, the various potentials
give more
or less the same pairing gap. Above this Fermi momentum, which
corresponds to a lab energy of $\sim 350$ MeV, the results start
to differ. This is simply due to the fact that the potentials
are basically fit to reproduce scattering data up to this
lab energy. Beyond this energy, the potentials predict rather
different phase shifts for the 
$^3P_2$-$^3F_2$ partial waves, see e.g.\  Ref.\ \cite{beehs98}.
\begin{table}[hbtp]
\begin{center}
\caption{Collection of $^3P_2$-$^3F_2$ energy gaps (in MeV) for the 
modern potentials discussed.  
BHF single-particle energies have been used. In case of no results,
a vanishing gap was found.}
\begin{tabular}{ccccc}\hline 
\multicolumn{1}{c}{$k_F\;({\rm fm}^{-1})$}& 
\multicolumn{1}{c}{CD-Bonn}&\multicolumn{1}{c}{$V_{18}$}&
\multicolumn{1}{c}{Nijm I}&\multicolumn{1}{c}{Nijm II} \\ \hline  
     1.2  & 0.04 & 0.04 & 0.04  & 0.04  \\
     1.4  & 0.10 & 0.10 & 0.10  & 0.10  \\
     1.6  & 0.18 & 0.17 & 0.18  & 0.18  \\
     1.8  & 0.25 & 0.23 & 0.26  & 0.26  \\
     2.0  & 0.29 & 0.22 & 0.34  & 0.36  \\
     2.2  & 0.29 & 0.16 & 0.40  & 0.47  \\
     2.4  & 0.27 & 0.07 & 0.46  & 0.67  \\
     2.6  & 0.21 &      & 0.47  & 0.99  \\
     2.8  & 0.17 &      & 0.49  & 1.74  \\
     3.0  & 0.11 &      & 0.43  & 3.14  \\ \hline
\end{tabular}
\label{tab:pgaps}
\end{center}
\end{table} 
Thus, before a precise calculation of $^3P_2$-$^3F_2$ energy gaps
can be made, one needs NN interactions that fit the scattering
data up to lab energies of $\sim 1$ GeV. This means 
in turn that the interaction models have to 
account for, due to the opening
of inelasticities above $350$ MeV, the
$N\Delta$ channel.

The reader should however note that the above results are
for pure neutron matter. We end therefore this section
with a discussion of the pairing gap for $\beta$-stable
matter of relevance for the neutron star cooling, see e.g.,
Ref.\ \cite{report}.
We will also omit a discussion on neutron pairing gaps in the
$^1S_0$ channel, since these appear at densities corresponding 
to the crust of the neutron star. The gap in the crustal material 
is unlikely
to have any significant effect on cooling processes \cite{pr95}, 
though
it is expected to be important in the explanation 
of glitch phenomena.
Therefore, the relevant pairing gaps for neutron star cooling
should stem from the 
the proton contaminant 
in the $^1S_0$ channel, and superfluid neutrons yielding energy gaps 
in the coupled $^3P_2$-$^3F_2$ two-neutron channel. 
If in addition one studies closely the phase shifts for
various higher partial waves of the NN interaction, one notices
that at the densities which will correspond to the  
core of the star, any superfluid 
phase should eventually disappear. This is due to the fact that
an attractive NN interaction is needed in order to
obtain a positive energy gap.

\begin{figure}\begin{center}
      {\epsfxsize=10pc \epsfbox{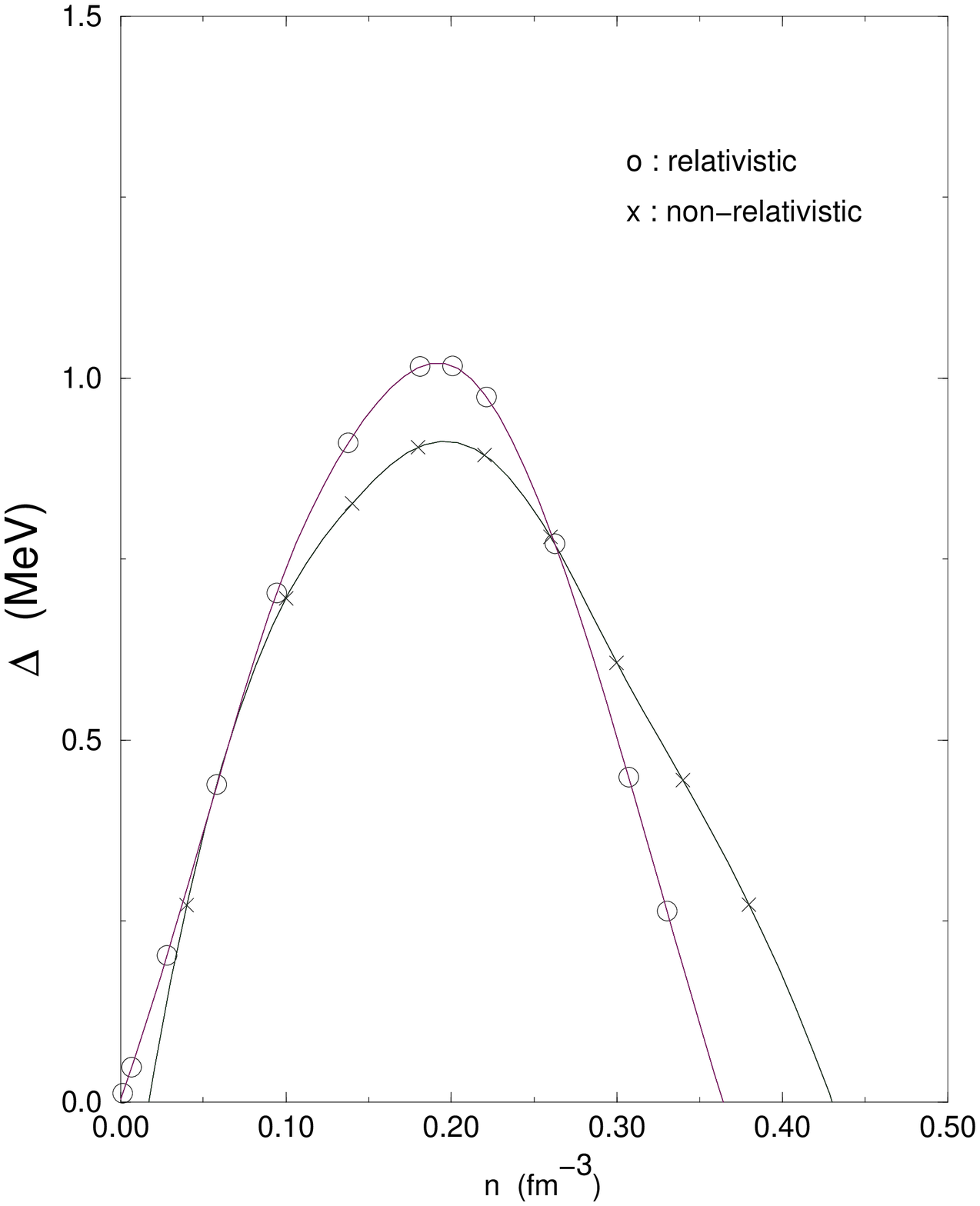}}\hspace{1cm}{\epsfxsize=10pc \epsfbox{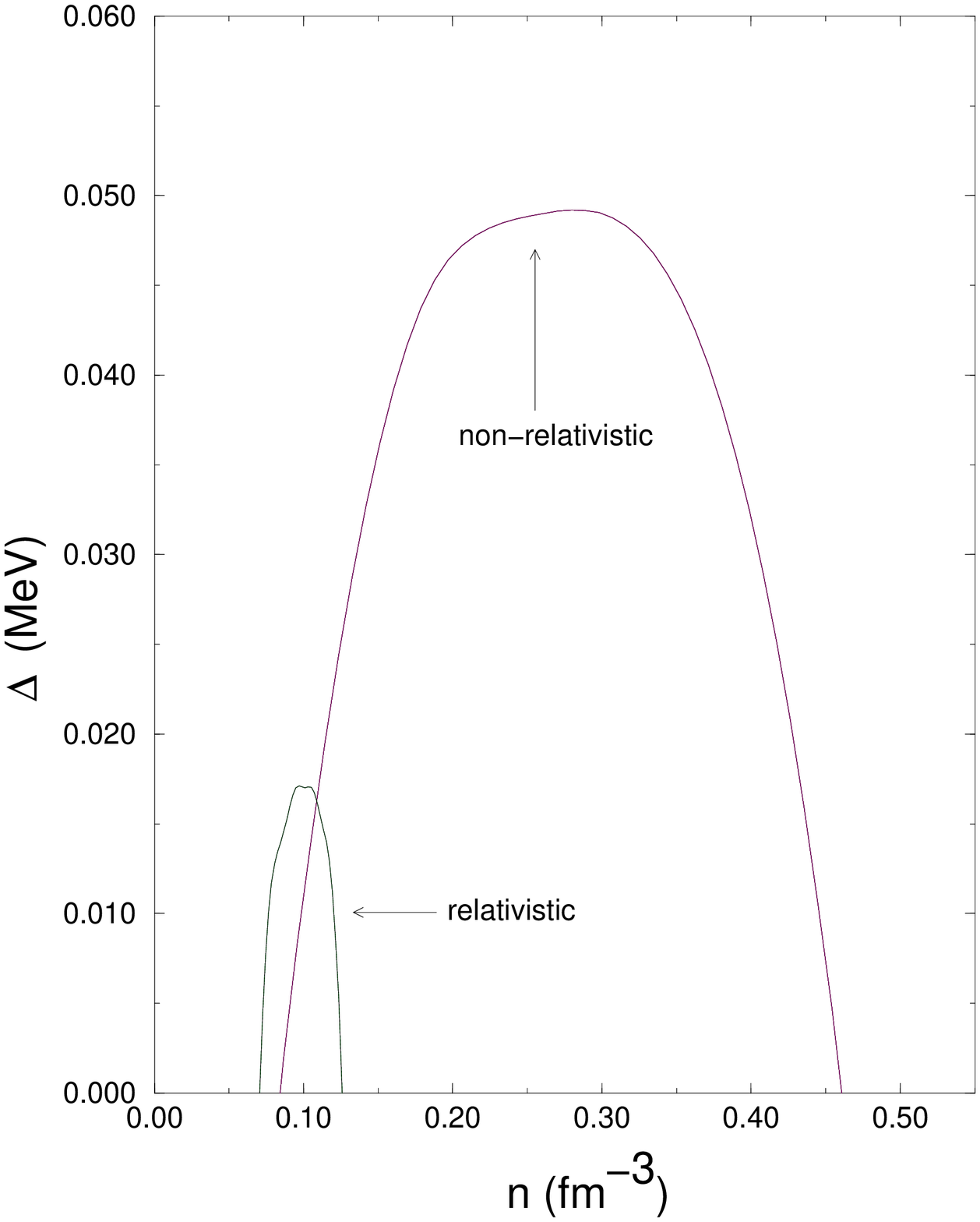}}
      \caption{Left box: Proton pairing in $\beta$-stable matter for 
          the $^1S_0$ partial wave. Right box: Neutron pairing in 
          $\beta$-stable matter for the $^3P_2$
          partial wave.}
     \label{fig:figgap}
\end{center}\end{figure}

Since the relevant total baryonic densities for these types of
pairing will be higher than the saturation
density of nuclear matter, we will account for relativistic
effects as well in the calculation of the pairing gaps.
As an example, consider the evaluation of the proton
$^1S_0$ pairing gap using a Dirac-Brueckner-Hartree-Fock  approach.
In Fig.\ \ref{fig:figgap} we plot as function of the total baryonic 
density the pairing gap for protons in the $^1S_0$
state, together with the results from the non-relativistic 
approach discussed in  Refs.\
\cite{elga96,eeho96}. 
These results are all 
for matter in $\beta$-equilibrium. In Fig.\ \ref{fig:figgap} 
we plot also the 
corresponding relativistic 
results for the neutron energy gap in the $^3P_2$ channel. 
For the 
$^3P_0$ and the $^1D_2$ channels we found both 
the non-relativistic and the relativistic
energy gaps to vanish. 

As can be seen from Fig.\ \ref{fig:figgap}, there are only small
differences (except for higher densities) between the non-relativistic
and relativistic proton gaps in the $^1S_0$ wave.
This is expected since the proton fractions (and their respective Fermi
momenta) are rather small.
For neutrons however, 
the Fermi momenta are larger, and we would 
expect relativistic effects to be important. At Fermi momenta
which correspond to the
saturation point of nuclear matter, $k_F=1.36$ fm$^{-1}$,
the lowest relativistic correction to the kinetic energy per 
particle is of the order of 2 MeV. 
At densities higher than the saturation
point, relativistic effects should be even 
more important.
Since we are dealing with
very small proton fractions, a Fermi momentum
of $k_F=1.36$ fm$^{-1}$, would correspond to a total baryonic 
density $\sim 0.09$  fm$^{-3}$. Thus, at larger densities 
relativistic effects for neutrons should
be important.
This is also reflected in Fig.\ \ref{fig:figgap} for the pairing
gap in the $^3P_2$ channel.
The relativistic $^3P_2$ gap is less  than half
the corresponding non-relativistic one, and the 
density region is also much smaller, see Ref.\ \cite{eeho96} for further details.

This discussion  can be summarized as follows.
\begin{itemize}
      \item The $^1S_0$ proton gap in $\beta$-stable matter
            is $ \le 1$ MeV, and if polarization
            effects were taken into account \cite{sclbl96},
            it could be further reduced by a factor 2-3.
      \item The $^3P_2$ gap is also small, of the order
            of $\sim 0.1$ MeV in $\beta$-stable matter.
            If relativistic effects are taken into account,
            it is almost vanishing. However, there is
            quite some uncertainty with the value for this
            pairing gap for densities above $\sim 0.3$
            fm$^{-3}$ due to the fact that the NN interactions
            are not fitted for the corresponding lab energies. 
      \item Higher partial waves give essentially vanishing
            pairing gaps in $\beta$-stable matter.
\end{itemize}
Thus, the $^1S_0$ and $^3P_2$ partial waves are crucial for our
understanding of superfluidity in neutron star matter.

\section{Finite nuclei} \label{sec:sec3}

We turn the attention to finite nuclei. Here we focus on the chain of 
tin isotopes. 
Considerable attention is at present being devoted to the
experimental and theoretical 
study of Sn isotopes \cite{schneider94,grawe95,fog94,hoff96,ehho95,ehho97}
ranging from the doubly-closed shell nuclei $^{100}$Sn to 
$^{132}$Sn.
Recently, substantial progress has been made in the 
spectroscopic approach to the neutron deficient doubly 
magic $^{100}$Sn core. Experimental
spectroscopic data are presently available down to $^{102}$Sn
\cite{grawe95}. 
Furthermore, detailed
spectroscopy for the other doubly closed-shell nucleus $^{132}$Sn
have recently been reported by Fogelberg {\em et al.}
\cite{fog94}. Similarly, the experimental single-particle
energies of $^{133}$Sn have recently been determined \cite{hoff96}.
The tin isotopes offer a unique opportunity for examining the
microscopic foundation of various phenomenological nuclear
models. In the Sn isotopes ranging from mass number $A=100$ to
$A=132$, neutrons are filling the subshells between the magic
numbers 50 and 82, and thus it is possible to examine how well
proton-shell closure at mass number 50 is holding up as valence
neutrons are being added, how collective features are developing,
the importance of certain many-body effects, etc.

Of interest in this study is the fact that 
the chain of even tin isotopes from $^{102}$Sn to $^{130}$Sn 
exhibits a near constancy of the 
$2^+_1-0^+_1$ excitation energy, a constancy which can be related
to strong pairing correlations and the near degeneracy in energy 
of the relevant single particle orbits. As an example, we show the 
experimental\footnote{We will limit our discussion to even nuclei
from  $^{116}$Sn to $^{130}$Sn, since a qualitatively similar picture
is obtained from $^{102}$Sn to $^{116}$Sn.}
$2^+_1-0^+_1$ excitation energy 
from  $^{116}$Sn to $^{130}$Sn in Table \ref{tab:table1}. 
Our aim is to see whether the partial waves which played such a crucial
role in neutron star matter, viz., $^1S_0$ and $^3P_2$, are equally
important in reproducing the near constant spacing in the chain
of even tin isotopes shown in  Table \ref{tab:table1}. 

To achieve this, we mount a large-scale shell-model calculation in 
a model space relevant for the description of tin isotopes. 
In order to test the dependence on the above partial waves in
the NN interaction, different effective interactions are employed.
 
Our scheme to obtain an effective two-body interaction for 
the tin isotopes
starts with a free nucleon-nucleon  interaction $V$ which is
appropriate for nuclear physics at low and intermediate energies. 
In this work we will thus choose to work with the charge-dependent
version of the Bonn potential models, see \mbox{Ref. \cite{cdbonn}}.
The next step 
in our many-body scheme is to handle 
the fact that the repulsive core of the nucleon-nucleon potential $V$
is unsuitable for perturbative approaches. This problem is overcome
by introducing the reaction matrix $G$ given by the solution of the
Bethe-Goldstone equation
\begin{equation}
    G=V+V\frac{Q}{\omega - H_0}G,
\end{equation}
where $\omega$ is the unperturbed energy of the interacting nucleons,
and $H_0$ is the unperturbed Hamiltonian. 
The operator $Q$, commonly referred to
as the Pauli operator, is a projection operator which prevents the
interacting nucleons from scattering into states occupied by other nucleons.
In diagrammatic language 
the $G$-matrix is the sum over all
ladder type of diagrams. This sum is meant to renormalize
the repulsive short-range part of the interaction. The physical interpretation
is that the particles must interact with each other an infinite number
of times in order to produce a finite interaction. 
We calculate $G$ using the double-partioning scheme discussed
in e.g.,~\mbox{Ref. \cite{hko95}}.
A harmonic-oscillator basis was chosen for the
single-particle
wave functions, with an oscillator energy $\hbar\Omega$ given
by
$\hbar\Omega = 45A^{-1/3} - 25A^{-2/3}=7.87 $ MeV,  $A=132$ being the mass
number.

Finally, we briefly sketch how to calculate an effective 
two-body interaction for the chosen model space
in terms of the $G$-matrix.  Since the $G$-matrix represents just
the summation to all orders of particle-particle
ladder diagrams, there are obviously other terms which need to be included
in an effective interaction. Long-range effects represented by 
core-polarization terms are also needed.
The first step then is to define the so-called $\hat{Q}$-box given by
\begin{equation}
   P\hat{Q}P=PGP +
   P\left(G\frac{Q}{\omega-H_{0}}G\\ + G
   \frac{Q}{\omega-H_{0}}G \frac{Q}{\omega-H_{0}}G +\dots\right)P.
   \label{eq:qbox}
\end{equation}
The $\hat{Q}$-box is made up of non-folded diagrams which are irreducible
and valence linked. The operator $P$ projects out the two-particle states defined
by the model-space.
Based on the $\hat{Q}$-box, we can in turn obtain 
an effective interaction
$H_{\mathrm{eff}}=\widetilde{H}_0+V_{\mathrm{eff}}^{(2)}$ in terms of the $\hat{Q}$-box,
 using the folded-diagram expansion, see e.g., Ref.\  \cite{hko95} for further details.

The effective two-particle matrix elements are calculated based on 
a $Z = 50, \quad N = 82$ asymmetric core and with the active $P$-space for holes
based on the $2s_{1/2}$, $1d_{5/2}$, $1d_{3/2}$, $0g_{7/2}$ and $0h_{11/2}$
hole orbits.
The corresponding single-hole energies are
$\varepsilon(d_{3/2}^{+}) = 0.00$~MeV, 
 $\varepsilon(h_{11/2}^{-}) = 0.242$~MeV, $\varepsilon(s_{1/2}^{+}) = 0.332$~MeV,
$\varepsilon(d_{5/2}^{+}) = 1.655$~MeV and  $\varepsilon(g_{7/2}^{+}) = 2.434$~MeV
and the shell model calculation amounts to studying
valence neutron holes outside this core.
The shell model problem requires the solution of a real symmetric
$n \times n$ matrix eigenvalue equation
\begin{equation}
       \widetilde{H}\left | \Psi_k\right\rangle  = 
       E_k \left | \Psi_k\right\rangle .
       \label{eq:shell_model}
\end{equation}
where for the present cases the dimension of the $P$-space reaches $n \approx 2 \times 10^{7}$.
At present our basic approach in finding solutions to Eq.(\ref{eq:shell_model})
is the Lanczos algorithm; an iterative method
which gives the solution of the lowest eigenstates. This method was 
already applied to nuclear physics problems by Whitehead {\sl et al.} 
in 1977. The technique is described in detail in Ref.\ \cite{whit77}, 
see also Ref.\ \cite{ehho95}.

\subsection{Different approaches to the shell-model effective interaction}

In order to test whether the $^1S_0$ and $^3P_2$ partial waves are equally
important in reproducing the near constant spacing in the chain
of even tin isotopes as they are for the superfluid properties of infinite matter,
we study four different approximations to the shell-model
effective interaction, viz.,
\begin{enumerate}
  \item Our best approach to the effective interaction, $V_{\mathrm{eff}}$, contains
        all one-body and two-body diagrams through third order in the $G$-matrix, 
        see Ref.\ \cite{ehho97}. 
  \item The effective interaction is given by the $G$-matrix only and inludes
        all partial waves up to $l=10$.
  \item We define an effective  interaction based on a $G$-matrix which now includes
        only the $^1S_0$ partial wave.
  \item Finally, we use an effective interaction based on a $G$-matrix which does
        not contain the  $^1S_0$ and $^3P_2$ partial waves, but all other waves
        up to $l=10$.  
\end{enumerate}
In all four cases the same NN interaction is used, viz., 
the CD-Bonn interaction described in Ref.\ \cite{cdbonn}.
Table \ref{tab:table1} lists the results obtained for the three first cases.  
\begin{table}[t]
\begin{center}
\caption{ $2^+_1-0^+_1$ excitation energy for the 
even tin isotopes $^{130-116}$Sn for various approaches
to the effective interaction. See text for further details. 
Energies are given in MeV. }\footnotesize
\begin{tabular}{lcccccccc}\hline
 & {$^{116}$Sn} & {$^{118}$Sn} & {$^{120}$Sn} &{$^{122}$Sn} & {$^{124}$Sn} & {$^{126}$Sn} & {$^{128}$Sn} & {$^{130}$Sn} \\ \hline
Expt & 1.29 & 1.23 & 1.17 & 1.14 & 1.13 & 1.14 & 1.17 & 1.23 \\
$V_{\mathrm{eff}}$ & 1.17 & 1.15 & 1.14 & 1.15 & 1.14 & 1.21 & 1.28 & 1.46 \\
$G$-matrix &1.14 & 1.12& 1.07 & 0.99 & 0.99 & 0.98 & 0.98 & 0.97  \\
$^1S_0$ $G$-matrix &1.38 &1.36 &1.34 &1.30 & 1.25& 1.21 &1.19 &1.18 \\\hline
\end{tabular}
\end{center}
\label{tab:table1}
\end{table}

We note from this Table that the three first cases nearly produce a constant 
$2^+_1-0^+_1$ excitation energy, with our most optimal effective interaction
$V_{\mathrm{eff}}$ being closest the experimental data. The bare $G$-matrix
interaction, with no folded diagrams as well, results in a slightly more compressed
spacing. This is mainly due to the omission of the core-polarization 
diagrams which typically render the $J=0$ matrix elements more attractive.
Such diagrams are included in $V_{\mathrm{eff}}$. 
Including only the $^1S_0$ partial wave in the construction of the  $G$-matrix
case 3,
yields in turn a somewhat larger spacing. This can again be understood from the
fact that a $G$-matrix constructed with this partial wave  
only does not receive contributions from any entirely repulsive partial wave.

It should be noted that our optimal interaction, as demonstrated in 
Ref.\ \cite{ehho97}, shows a rather good reproduction of the 
experimental spectra for both even and odd nuclei. Although the approximations
made in cases 2 and 3 produce an almost constant $2^+_1-0^+_1$ excitation energy,
they reproduce poorly the properties of odd nuclei and other 
excited states in the even Sn isotopes. 

However, the fact that the first three  approximations result in a such a good
reproduction of the  $2^+_1-0^+_1$ spacing may hint to the fact that the 
$^1S_0$ partial wave is of paramount importance. 
If we now turn the attention to case 4, i.e., we omit the
$^1S_0$ and $^3P_2$ partial waves in the construction of the $G$-matrix,
the results presented  in Table \ref{tab:table2} exhibit  a spectroscopic 
catastrophe. In this Table we do also not list eigenstates
with other quantum numbers. For e.g., $^{126}$Sn
the ground state is no longer a $0^+$ state, rather it carries the quantum numbers
$4^+$ while for $^{124}$Sn the ground state 
has $6^+$. The first $0^+$ state for this nucleus is given at an excitation
energy of $0.1$ MeV with respect to the $6^+$ ground state.
The general picture for other eigenstates is that of an extremely poor agreement
with data.  
\begin{table}[t]
\begin{center}
\caption{ $2^+_1-0^+_1$ excitation energy for the 
even tin isotopes $^{130-124}$Sn obtained with a $G$-matrix
effective interaction which excludes the important
pairing waves $^1S_0$ and $^3P_2$. See text for further details. 
Energies are given in MeV. }
\begin{tabular}{lcccc}\hline
& {$^{124}$Sn} & {$^{126}$Sn} & {$^{128}$Sn} & {$^{130}$Sn} \\ \hline
No $^1S_0$ and $^3P_2$ in $G$-matrix &0.15  &-0.32  &0.02 &-0.21  \\\hline
\end{tabular}
\end{center}
\label{tab:table2}
\end{table}
Since the agreement is so poor, even the qualitative reproduction of the 
$2^+_1-0^+_1$ spacing, we defer from performing time-consuming shell-model
calculations for $^{116,118,120,122}$Sn.

\section{Conclusion} \label{sec:sec4}

In summary, the $^1S_0$ and $^3P_2$ partial waves are crucial for our
understanding of superfluidity in neutron star matter. In addition, viewing
the results of Table \ref{tab:table2}, one sees that 
pairing correlations, being important for the 
reproduction of the $2^+_1-0^+_1$ excitation energy of
the even Sn isotopes, depend strongly on the same  partial waves
of the NN interaction. Omitting these waves, especially the  $^1S_0$ wave,
results in a spectrum which has essentially no correspondence with experiment.
Further analyses of the results for finite nuclei will be presented elsewhere
\cite{eho2000}.

\end{document}